\documentclass[10pt,conference]{IEEEtran}
\IEEEoverridecommandlockouts

\usepackage{times}
\usepackage{graphicx}
\usepackage{xspace,shortcut}
\usepackage{comment}
\usepackage{color,soul}
\usepackage{tcolorbox}
\tcbuselibrary{skins}
\usepackage{listings}

\def\BibTeX{{\rm B\kern-.05em{\sc i\kern-.025em b}\kern-.08em
    T\kern-.1667em\lower.7ex\hbox{E}\kern-.125emX}}

\definecolor{backcolour}{rgb}{0.95,0.95,0.92}

\begin{document}
\title{\texttt{BeeSwarm}: Enabling Scalability Tests in Continuous Integration}

\author{
\IEEEauthorblockN{
Jieyang Chen,
Qiang Guan\IEEEauthorrefmark{1},
Li-Ta Lo\IEEEauthorrefmark{2},
Patricia Grubel\IEEEauthorrefmark{2}, 
and Tim Randles\IEEEauthorrefmark{2},
}

\IEEEauthorblockA{
Oak Ridge National Laboratory, Oak Ridge, TN, USA\\
}

\IEEEauthorblockA{\IEEEauthorrefmark{1}Kent State University, Kent, OH, USA\\
}
\IEEEauthorblockA{\IEEEauthorrefmark{2}Los Alamos National Laboratory, Los Alamos, NM, USA
}
chenj3@ornl.gov qguan@kent.edu \\
\{ollie, pagrubel, trandles\}@lanl.gov
}



\maketitle

\linespread{1.1}%
\selectfont

\begin{abstract}
Testing is one of the most important steps in software development. It ensures the quality of software. Continuous Integration (CI) is a widely used testing system that can report software quality to the developer in a timely manner during the development progress. Performance, especially scalability, is another key factor for High Performance Computing (HPC) applications. Though there are many applications and tools to profile the performance of HPC applications, none of them are integrated into the continuous integration. On the other hand, no current continuous integration tools provide easy-to-use scalability test capabilities.
In this work, we propose BeeSwarm, a scalability test system that can be easily applied to the current CI test environment enabling scalability test capability for HPC developers. As a showcase, BeeSwarm is integrated into Travis CI and GitLab CI to execute the scalability test workflow on Chameleon cloud. 

\end{abstract}

\begin{IEEEkeywords}
scalability test; continuous integration; high performance computing; cloud computing; container.
\end{IEEEkeywords}



\footnotetext[1]{The publication has been assigned the LANL identifier LA-UR-18-25223.}
\section{Introduction}
High software quality is one of the most important goals of software development. Software testing serves as the most widely used approach to ensure the quality of software meet expectation.
A good way to test software is to include automated tests in the build process.  With the rise of Extreme Programming (XP) and Test Driven Development (TDD), self-testing processes for code development have become popular and are widely adopted by many software development projects. 
As software becomes increasingly structurally complicated, the number of developers involved in the development process increases. As each developer makes progress, they commit their work periodically (every several hours or days) to the central code repository (e.g., git, SVN). Not only does each developer's work require testing, the integration of work between developers also requires testing. So, Continuous Integration (CI) \cite{fowler2006continuous} is widely adopted in many software development projects. A CI server is used dedicatedly for testing. Each time a developer makes a commit of her work to the central code repository, the CI server automatically make a clone of the project and conduct pre-designed tests, so that it can constantly monitor the quality of the software in terms of correctness and report potential problems in a timely fashion, helping developers make bug fixes more efficiently.  

\begin{figure*}[h]
    \centering
    \includegraphics[width=1\textwidth]{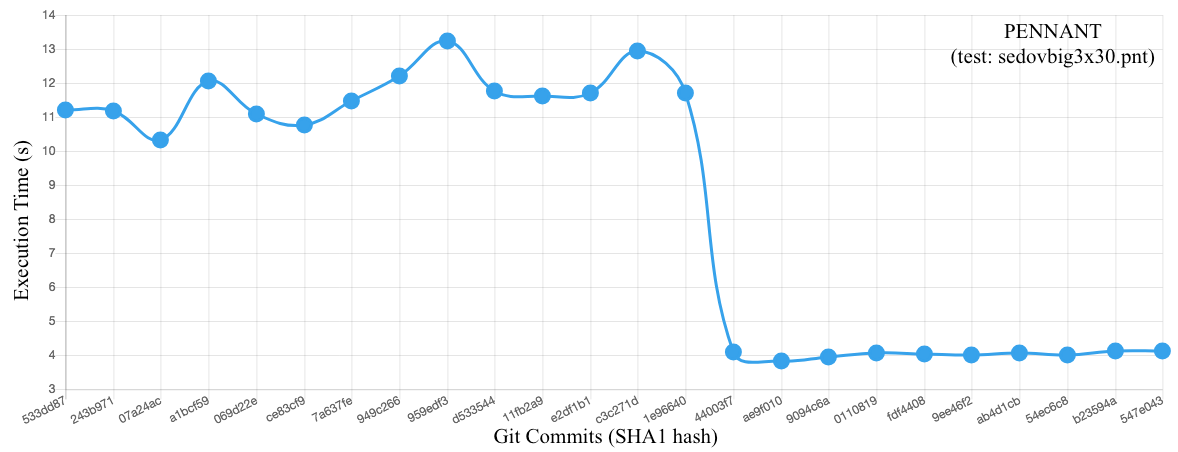}
    \caption{Example: the performance of Legion \cite{bauer2012legion} changes as developers make progress. The performance is obtained by running a benchmark PENNANT\cite{ferenbaugh2015pennant} on the Legion system. The test suit sedovbig3x30 running on 10 processes (CPU cores) is used. 
}
    \label{exp}
\end{figure*}

When it comes to HPC applications, \textit{performance} and \textit{scalability} 
are the other two important factors of software quality besides correctness, since the application are usually designed to deliver high performance on given platforms. Also applications that aim to solve complex time-consuming problems are expected to obtain good speedup when deployed on multi-node clusters, many-core architectures, or large-scale supercomputers. The scalability of HPC application is usually interpreted as how much speed up can be obtained given more computing resources. Better scalability means that the HPC application can use the underlying computing resources more efficiency and constantly deliver good performance on a various amount of computing resources.

During the HPC application development, as developers make progress,
and they commit their work to the central code repository, the scalability of the application can change. For instance, it can be caused by changes in algorithm design, tunable parameters, and different hardware architectures of target production systems. For example, \textbf{Fig. \ref{exp}} shows the  performance of Legion \cite{bauer2012legion}, a data-centric parallel programming system, changes 
with different source code commits. The performance is obtained by running a benchmark software, PENNANT\cite{ferenbaugh2015pennant}, on the Legion system. As we can see the execution time can significantly change as developers make progress. Receiving performance or scalability results like this in a timely manner can greatly help developers make better decisions about their code design and deliver HPC software with expected quality. However, current designs of CI services are commonly focused on monitoring the software quality in terms of correctness (e.g., detecting software bugs). To the best of our knowledge, none of the current work can easily enable automatic performance or scalability tests in CI since the test environment of CI is usually deployed on a single machine incapable of conducting large-scale scalability test.

\begin{figure*}[h]
    \centering
    \includegraphics[width=0.8\textwidth]{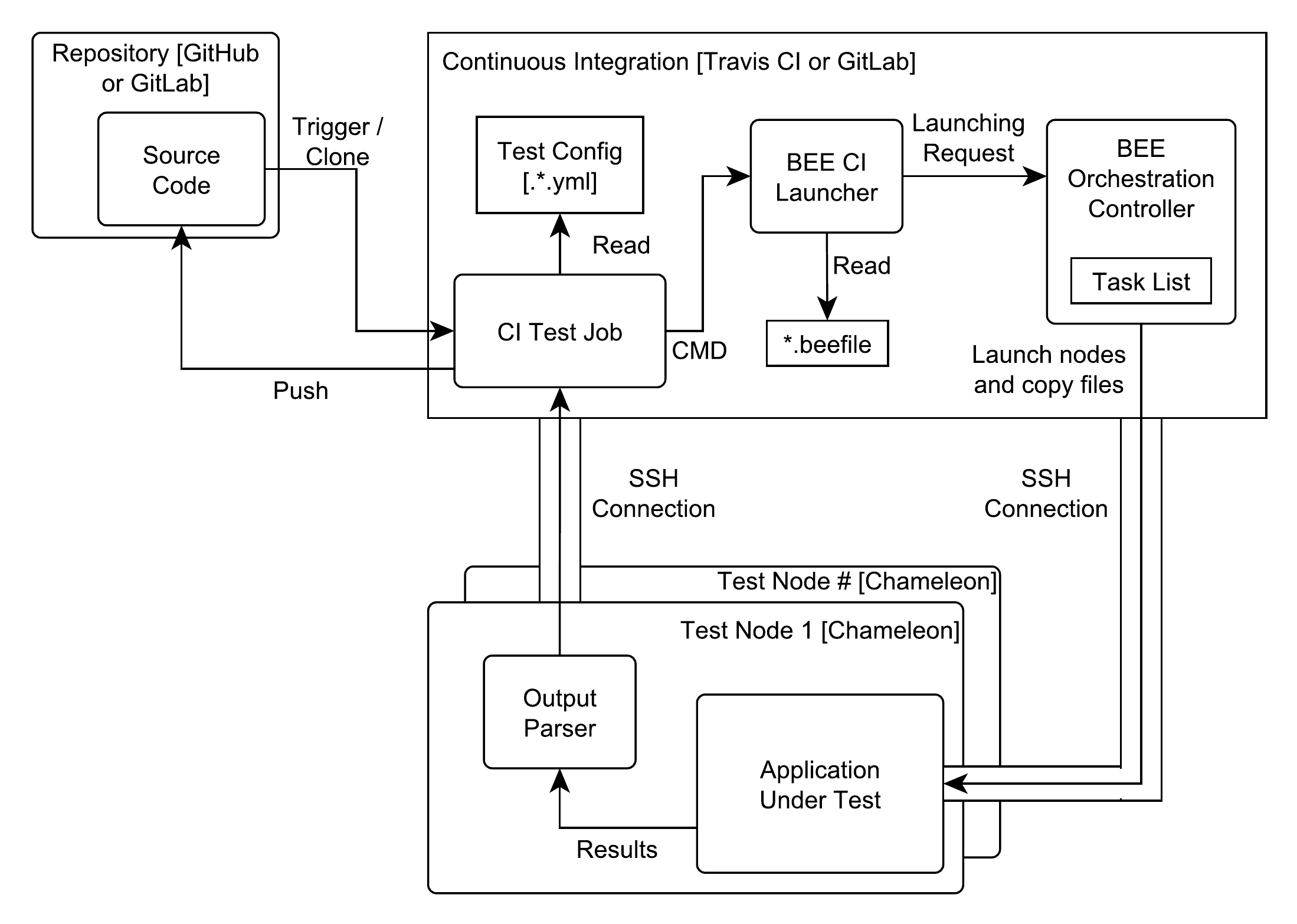}
    \caption{Architecture of \texttt{BeeSwarm}. 
    }
    \label{arch}
\end{figure*}

In this work, we propose a performance and scalability test system for CI -- \texttt{BeeSwarm}. \texttt{BeeSwarm} can be used as a plug-in for any current CI service. It takes the widely used Docker container as input, and the performance and scalability test can run on both HPC cluster environments and cloud computing environments. Just like the original correctness test in CI, the performance and scalability test are also autonomic. It only requires users to make simple specifications about the test environment they want to use and the test specification they need. Every time developers commit a change to the central code repository, they can choose to schedule a scalability test after the success of original correctness test. The performance and scalability test results will be automatically pushed back to the central code repository. 
Although we deploy \texttt{BeeSwarm} on Travis CI and GitLab CI in this work, it can also be deployed on any other CI test environment. To deploy on another CI platform, only minimum modifications to the \texttt{BeeSwarm} configuration scripts are necessary, which makes \texttt{BeeSwarm} highly portable across CI platforms. In addition, although we only show the use of Chameleon cloud, the scalability test can also be executed on any other BEE-supported platform (HPC clusters, AWS, OpenStack, etc). This gives developers the flexibility to choose the platform they want their applications to run on.

The rest of this paper is organized as follows. We motivate our work in section \ref{motivation}. In section \ref{background}, we give necessary background that can help readers understand this work. We provide design details of \texttt{BeeSwarm} in section \ref{design} followed by experimental evaluation in section \ref{experiments}. Section \ref{related_work} discuss recent work that related to ours. Finally, section \ref{conclusion} concludes our work. 

\section{Motivation}
\label{motivation}

\begin{table}[h]
\centering
\caption{Several commits in the Legion commit tree that may cause the performance improvement after commit 4400 shown in Figure \ref{exp}.}
\label{commits}
\begin{tabular}{|p{0.3cm}|p{5.8cm}|}
\hline
\multicolumn{1}{|c|}{Commit HASH} & \multicolumn{1}{c|}{Commit message}                                                                                                  \\ \hline
725e549dc                         & legion: fixing a potential hang with old-style bounds checking                                                                       \\ \hline
3edff3290                         & regent: small bug fix to openmp code generation for regent                                                                           \\ \hline
d0b157755                         & tools: small bug fix for legion prof ascii deserializer                                                                              \\ \hline
1162649ea                         & legion: small bug fix for dependence analysis of close operations involving different children in different modes for the same field \\ \hline
2818b5fe9                         & legion: small bug fix for remote advances of version numbers                                                                         \\ \hline
824d6c77d                         & legion: fixing a bug where we were not properly paging in version states for remote virtual mappings                                 \\ \hline
\end{tabular}
\end{table}

In this section, we use an example to motivate our work by showing the necessity of having automatic scalability test in CI. In \textbf{Fig. \ref{exp}} we show the performance of Legion changes as developers make progress. However,  it is hard to find out the exactly which commit(s) causes the performance change. For example, the performance of Legion improved significantly from commit \texttt{1e96} to \texttt{4400}. Commit \texttt{4400} is a merge operation between two branches, which totally contains about 61300 lines of code changes composing hundreds of commits. It is hard to tell which commit(s) causes the performance improvement. By searching the commit tree of Legion, we found several commits focusing on bug fixing that may potentially affect performance. We list several of them in \textbf{table \ref{commits}}. So, if scalability test was available in the CI for Legion upgrade, we would be able to easily find the root cause of the performance change by searching in the scalability test results for each commit and keep track of the changes that benefit or hurt the scalability. 


\section{Background}
\label{background}
\subsection{Build and Execution Environments (BEE)}
\texttt{BEE} \cite{bee,beeflow,chen2018build} is a 
containerization environment that enables HPC applications to run on both HPC and cloud computing platforms. \texttt{BEE} provides a unified user interface for automatic job launching and monitoring. \texttt{BEE} users only need to wrap their applications in a standard Docker image and provide a simple \texttt{BeeFile} (job execution environment description) to run on \texttt{BEE}. Since the same Docker image is used 
 across platforms, no source code modification is necessary. 
In this work, we build \texttt{BeeSwarm} based on \texttt{BEE}, so it naturally inherits all benefits of \texttt{BEE}. This allows us to build a unified scalability test system across multiple platforms.

\subsection{Continuous Integration (CI)}
CI was first named and proposed by Grady Booch in 1991. Its aim was to greatly reduce integration problems. CI was initially combined with automated unit testing to run on the developer's local machine before committing to the central code repository.  However, as software being developed becomes more complicated and more people are involved in developing, localized testing becomes inefficient and the code base on each developer's machine can easily become outdated, so integration can still be problematic. The longer a branch of code remains checked out, the greater the risk of multiple integration conflicts and failures when the developer branch is reintegrated into the main line. So, centralized build servers are used for CI. The build servers can perform more frequent (e.g., every commit) test runs and provide reports back to the developers. Driven by these benefits many HPC application development projects are now using CI. For example, almost all projects in Next-Generation Code Project in Los Alamos National Laboratory are using CI \cite{daniel2016lanl}. Currently, many CI tools are available to developers such as Travis CI, GitLab CI, Circle CI, Codeship, etc. Many computing platforms also provide CI as a feature in their services such as AWS, Azure, etc. However, current designs of CI services only focus on detecting software bugs in the HPC softwares. To the best of our knowledge, none of the current work can easily enable automatic scalability tests in CI. So, in this work we propose to enable easy scalability tests for HPC developers. 

\begin{figure*}[h]
    \centering
    \includegraphics[width=1\textwidth]{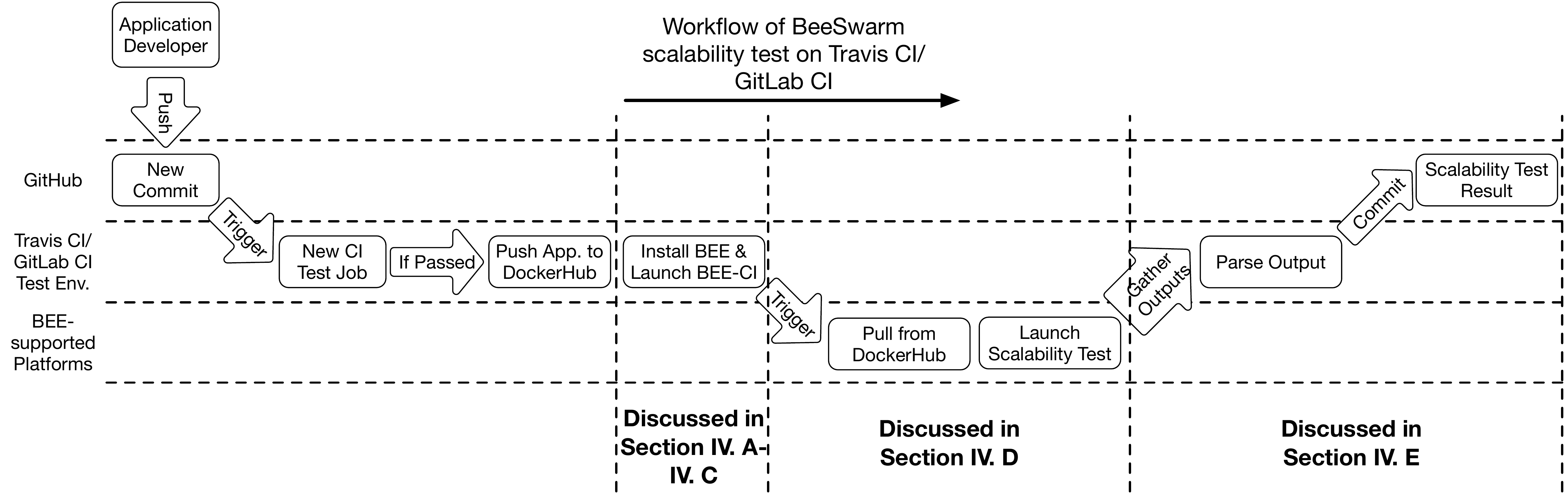}
    \caption{Overall Workflow of BeeSwarm CI Scalability Test. 
    }
    \label{overall}
\end{figure*}

\section{Design}
\label{design}
In order to fulfill the goals of \texttt{BeeSwarm} the software architecture required would require that we both leverage industry standards, while at the same time implementing new functionality to \texttt{BEE}\cite{bee}. \texttt{BeeSwarm} is a general solution that can be deployed on any git repository, any CI service and any BEE-supported computing platform. For the purposes of example the software platforms 
\texttt{Travis CI} and \texttt{GitLab CI} are used as two CI platform, and \texttt{Chameleon Cloud}\cite{mambretti2015next}
are used a scalability test platform. \textbf{Fig. \ref{arch}} shows the architecture of \texttt{BeeSwarm}. \texttt{BEE} is at the core of the architecture and serves a number of vital roles. As part of the continuous integration process \texttt{BEE} is deployed on the CI test environment, from there it is responsible for managing the workflow associated with creating a scalable test environment, copying required test scripts, initiating the target application, and finally parsing the output. 
\textbf{Fig. \ref{overall}} shows the workflow of \texttt{Travis CI}/\texttt{GitLab CI} with \texttt{BeeSwarm}. Once developers make commits to the central code repository, the original CI correctness test will be triggered. If the test finishes without a fail, \texttt{BeeSwarm} will start to deploy the scalability on BEE-supported computing platform, gather the results and push back to the code repository. It is crucial that we use \texttt{BeeSwarm} to conduct scalability test, since the CI test environment is usually deployed on a single machine incapable for large-scale scalability test. There are four major design tasks in \texttt{BeeSwarm} and we discuss them as follows.



\subsection{Integrate BEE in CI Test Environment}
Each time a developer commits to the central code repository, a new CI test job is triggered on the CI test environment. That means in order to launch BEE inside that test environment, we need to install BEE every time before the scalability test. To minimize overhead caused by the installation, we designed a more efficient customized BEE installer for CI environment. Since BEE does not run any test locally in the CI environment, we remove the image building process that was originally in the BEE installer, required when BEE runs jobs in a virtual machine on a system. 
Also, we design a simplified BEE launcher (discussed in the next subsection), which requires less dependent packages/libraries, simplifying the BEE installer. Finally, to enable remote control of compute platforms through SSH, we add SSH key generation in the new BEE installer. This was not present in the original BEE installer, since it can utilize the current user's key. With all kinds of optimization, we are able to keep the BEE installation time to less than two minutes, causing only a slight overhead compared with minutes to hours of CI test and scalability test.

\subsection{Customize BEE Launcher for CI Test Environment}
BEE was designed to handle multiple tasks simultaneously, so it adopted a server-client structure, in which the server is a centralized controller (i.e., \texttt{BEE Orchestration Controller}) that stores the global information of all running BEE jobs and clients, a series of BEE launchers (each targeting a computing platform). This structure can facilitate normal use, however it can be cumbersome to launch BEE jobs on a CI test environment using the server-client structure (first start the \texttt{BEE Orchestration Controller} in the background and then launch the job using the BEE launcher).  Since we only run one BEE job for each CI job, there is no need to use the centralized controller to keep all the information of multiple jobs. So, in this work we design a simplified BEE launcher. It allows Travis to launch the BEE job with just one simple command. Basically, we integrate the input parser and job launching process together in our simplified BEE launcher.

\subsection{Customized \texttt{beefile} }
\texttt{beefile} is a simple JSON-format task description file used by BEE as user input. It contains necessary information needed to launch a task using BEE that include Docker images tag, platform-specific settings, and run script for both sequential runs and parallel runs. Here, we extend the run script configuration part for parallel runs. In the original design, uses need to specify each parallel run command one by one including the script to invoke, number of node to use, and number of processes to be used per node. Since users usually only need to run a few parallel run command, this design is clear and simple to use. However, for scalability test, users expect to run their application with a series of configurations (e.g., increasing number of nodes/processes). Fill in each configuration one by one can be cumbersome. So, we extend the \texttt{beefile} to allow easier configuration. Specifically, instead of letting users specify each configuration one by one, we now allow users to specify a range of configurations. For example, a range of nodes and a range of processes per node. In addition, we also users to specify, whether they want to increase the number of nodes or processes linearly with a fixed step size or logarithm with base of two. An example \texttt{beefile} is shown in \textbf{Listing 1}.

\lstset{numbers=left,
xleftmargin=1.5em,
frame=single,
framexleftmargin=3em}

\lstset{language=Java}
\begin{lstlisting}[ float, escapechar=!,
				   caption= An example \texttt{beefile}]
 "task_conf": {
  "task_name": <task name>,
  "exec_target": bee_cc|bee_vm|bee_aws|bee_os,
  !\hl{"scalability\_test": \{}!
    !\hl{"script" : \<path to script\>,}!
    !\hl{"num\_of\_nodes": [1, 32],}!
    !\hl{"proc\_per\_node": [1, 16],}!
    !\hl{"mode": linear or log}!
  !\hl{\},}!
 "docker_conf": {
  "docker_img_tag": <docker image>,
  "docker_username": <username>,
  "docker_shared_dir": <dir>
 },
 "exec_env_conf": {
  "bee_cc": {...} or
  "bee_vm": {...} or
  "bee_aws": {...} or
  "bee_os": {...} 
 }       

\end{lstlisting}

\subsection{Test Scalability  on BEE-supported Platform}
Since CI services usually only allocate one computing node (e.g., virtual machine) for each job, it is impractical to conduct a scalability test beyond one node. So, in this work we choose to use BEE as the computing back-end for the scalability test. BEE supports launching any kind of computing task on a variety of computing platforms, ranging from HPC systems to cloud computing systems (e.g., Amazon EC2, OpenStack). It can launch each job on as many nodes as each computing platform allows. BEE takes a job description file, \texttt{Beefile}, as input, that specifies all job related information including selecting the target platform, name tag of the Docker container for the application, and run scripts that the user specifies to be run when the application is deployed on the target platform. To launch a BEE job for the scalability test, we keep using the same \texttt{Beefile} as the job description. To specify specific test configurations for the scalability test, users only need to add multiple entries to the "mpirun" section inside the \texttt{Beefile}. Deploying the execution environment on the target system can take several minutes, to avoid setting up the environment repeatedly for each test, the BEE-CI launcher will first scan through the \texttt{Beefile}, and then setup the environment with the maximum number of nodes needed to conduct all tests. 

\subsection{Collect and Store Scalability Test Results}
Unlike common CI tests that only provide results in the form of ``\textit{pass}'' or ``\textit{no pass}'' to developers, scalability test reports a variety of information generated from different execution scales to developers. Since the information that developers care about is different from application to application, it is hard to develop a universal monitor strategy to gather information that suits everyone's needs. So instead, we leave this part to the developers. We let developers program their applications so that after each run the application will output relevant information. BEE will gather all the outputs from different runs as separate files, transferred and saved in the CI test environment. Next, we require developers to provide an output parser that can parse all relevant information from the output files and generate one final result file. Finally, BEE will push the final result file back to the central code repository and rename the file using the git build number to distinguish final result files generated from different commits. 
\section{Experiments}
\label{experiments}
In this section, we conduct experiments to show the performance and scalability of \texttt{BeeSwarm}. We use a Department of Energy (DOE) code, FleCSALE \cite{charest2017flexible}, as an example software development project. FleCSALE is a computer software package developed for studying problems that can be characterized using continuum dynamics, such as fluid flow. It is specifically developed for existing and emerging large distributed memory system architectures. We deploy \texttt{BeeSwarm} on both Travis CI and GitLab CI. For Travis CI, We use the default virtual machine based execution environment to run the original correctness test and \texttt{BeeSwarm}. For GitLab CI, we user the Docker-in-Docker (i.e., dind) runner to run the original correctness test and \texttt{BeeSwarm}. We found that Docker-in-Docker runner enables an more easy-to-configured environment for \texttt{BeeSwarm} compared to other runner types.  We use Chameleon Cloud \cite{mambretti2015next} as the computation back-end for the scalability tests. The Chameleon Cloud is an OpenStack-based cloud computing platform that offer bare-metal access to all computing nodes. It is currently deployed at University of Chicago and the Texas Advanced Computing Center with total 650 multi-core nodes. We conduct our test on the nodes located at University of Chicago. 

\subsection{Modified CI script}
In this section, we show a sample modified Travis CI script (similar on GitLab CI) for FleCSALE that has \texttt{BeeSwarm} scalability test enabled (\textbf{Listing 2}). Line 1 - 13 are the original FleCSALE test code on Travis CI. To enable \texttt{BeeSwarm} scalability test, we only need to add less than 10 lines of simple code (line 14 - 23). The original CI script include building a Docker image  (line 9), running the Docker image (line 10) to correctness test scrips, and push the image to DockerHub if the test was successful (line 14). We add the \texttt{BeeSwarm} configuration and launching scripts after the image is successfully pushed onto the DockerHub. We obtain and install \texttt{BeeSwarm} in line 14 - 16. We add necessary environment variables (for OpenStack and \texttt{BeeSwarm}) in line 17. The scalability test is launched using a simple command in line 18. We add a 120 minutes timeout here since Travis CI would kill a job if a command runs more than 10 minutes by default and a scalability test usually needs more time than that. The actual timeout length can be set based on need of a specific application. Finally, we run the output parser in line 19 followed by pushing scalability test result to original code repository in line 20 - 23. It can be seen that with minimum modification current CI scripts can easily enable scalability test through \texttt{BeeSwarm} and the scalability test code is highly portable across any kind of CI service platforms.

\lstset{numbers=left,
xleftmargin=1.5em,
frame=single,
framexleftmargin=2em}

\lstset{language=Java}
\begin{lstlisting}[float, escapechar=!,
				   caption= Example Travis CI script (\texttt{.travis.yml}) for FleCSALE with \texttt{BeeSwarn} scalability test. Highlighted part shows that only simple modifications are required to enable autonomic scalability test.]
language: cpp
sudo: required
services:
 - docker
before_install:
 - git fetch --unshallow 
 - git fetch --tags
script:
 - docker build -t <img> <dockerfile> 
 - docker run <img> <correctness_test>
after_success:
 - docker login -u $DOCKER_USERNAME -p $DOCKER_PASSWORD
 - docker push <img>
 - !\hl{git clone https://github.com/lanl/BEE.git}!
 - !\hl{cd ./BEE}!
 - !\hl{./install\_on\_travis.sh}!
 - !\hl{source openrc.sh}!
 - !\hl{travis\_wait 120 bee\_ci\_launcher.py -l FleCSALE}!
 - !\hl{output\_parser.py}!
 - !\hl{git add scalability\_test\_result\_\$BUILD\_NUM.csv}!
 - !\hl{git commit --message "BeeSwarm commit \$BUILD\_NUM [skip ci]"}!
 - !\hl{git remote add remote\_repo https://\$REPO\_TOKEN@\$REPO\_URL }!
 - !\hl{git push --quiet --set-upstream remote\_repo \$BRANCH}!

\end{lstlisting}

\subsection{Required environment variables}
\begin{table}[h]
\caption{List of variables needed by \texttt{BeeSwarm} in CI environment}
\label{var}
\begin{tabular}{|p{3cm}|p{5cm}|}
\hline
Variable & Description \\ \hline
DOCKER\_USERNAME &  Username for Docker image registry.\\ \hline
DOCKER\_PASSWORD &  Password for Docker image registry.\\ \hline
REPO\_TOKEN & Access token used for pushing scalability test results back to the code repository. \\ \hline
REPO\_URL & The URL to the code repository. \\ \hline
REPO\_BRANCH  & The current branch of the code repository. \\ \hline
BUILD\_NUM & Current build number. \\ \hline
OS\_USERNAME & Username for accessing OpenStack platform. \\ \hline
OS\_PASSWORD & Password for accessing OpenStack platform. \\ \hline
OS\_RESERVATION\_ID & Reservation ID used for current scalability test on OpenStack platform. \\ \hline
\end{tabular}
\end{table}

\textbf{Table \ref{var}} lists the variables that are necessary for \texttt{BeeSwarm} in the CI test environment. \texttt{DOCKER\_USERNAME} and \texttt{DOCKER\_PASSWORD} are used to access (e.g., pull and push) Docker images from the images registry. \texttt{REPO\_TOKEN} is used to let \texttt{BeeSwarm} push the scalability test results back to the original code repository. \texttt{REPO\_BRANCH} and \texttt{BUILD\_NUM} are used to make sure that \texttt{BeeSwarm} will push the scalability test results back to the corresponding branch with build number marked in the commit message. \texttt{OS\_USERNAME} and \texttt{OS\_PASSWORD} are used to access the OpenStack platforms (e.g., Chameleon cloud) and \texttt{OS\_RESERVATION\_ID} is used to specify a list of nodes used for scalability test. 

\subsection{Performance of \texttt{BeeSwarm}}
In order to evaluate the performance of \texttt{BeeSwarm}, we discuss the overhead of launching \texttt{BeeSwarm} and the scalability of \texttt{BeeSwarm} for large-scaled test.

\subsubsection{Overhead of \texttt{BeeSwarm}}
\begin{figure}[h]
    \centering
    \includegraphics[width=0.48\textwidth]{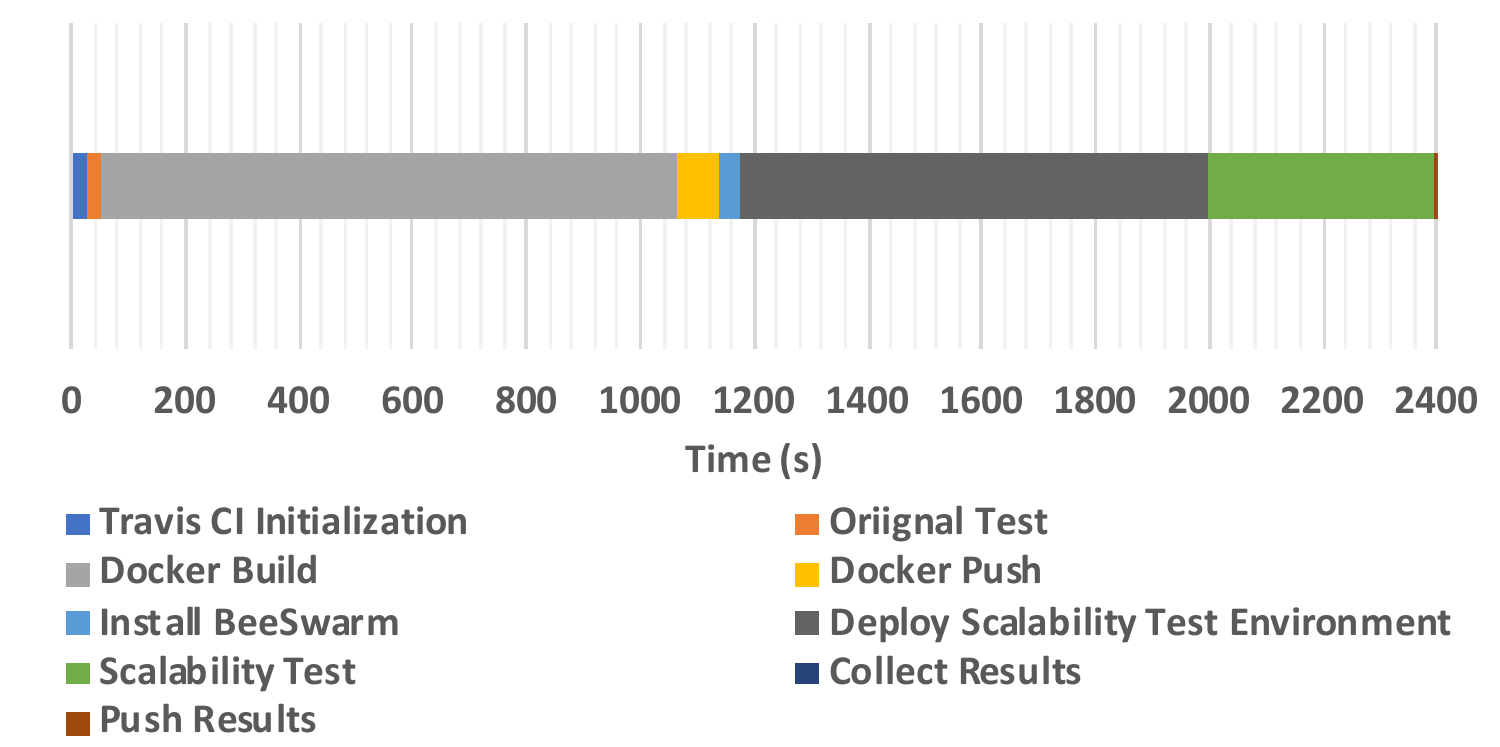}
    \caption{Time breakdown of an example CI test with \texttt{BeeSwarm} scalability test on Travis CI.}
    \label{breakdown}
\end{figure}

\begin{figure}[h]
    \centering
    \includegraphics[width=0.48\textwidth]{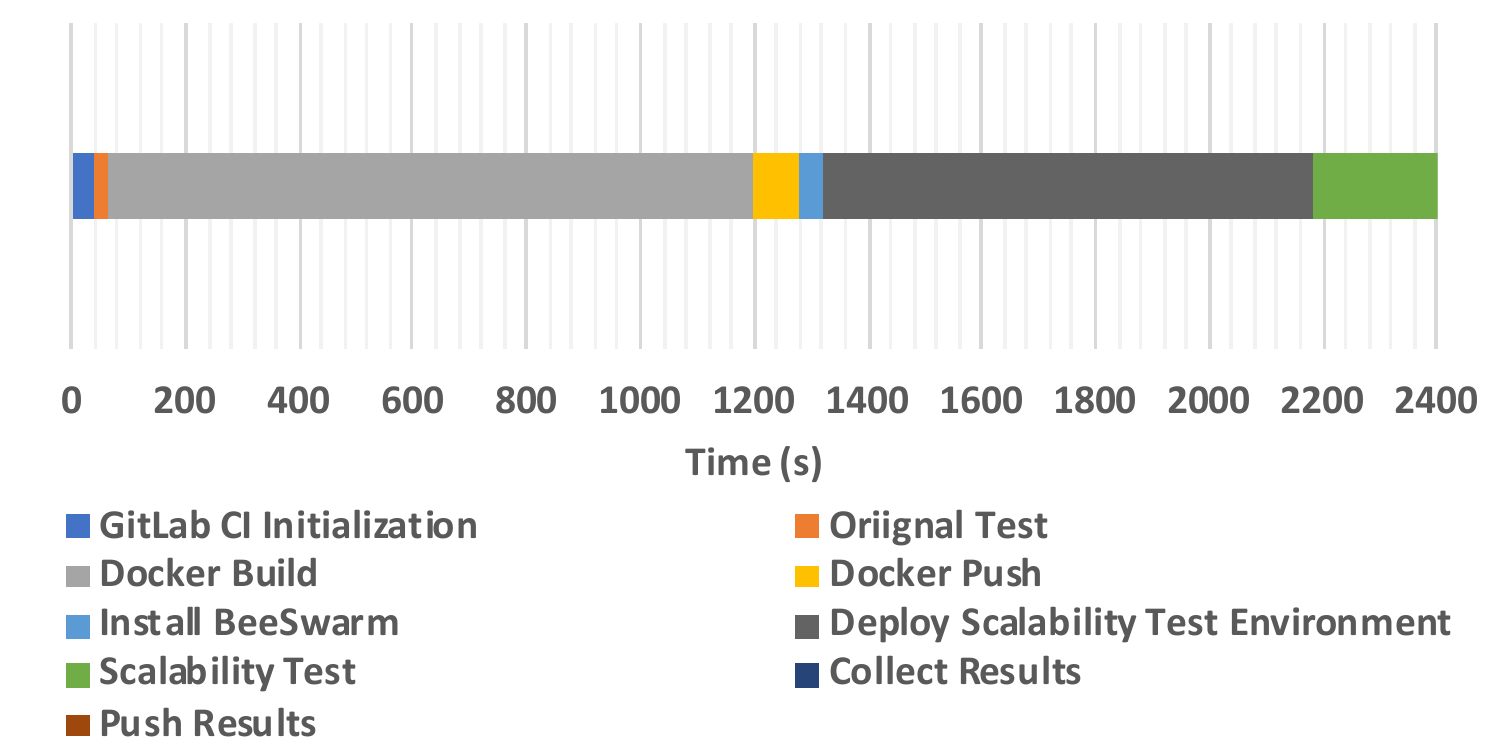}
    \caption{Time breakdown of an example CI test with \texttt{BeeSwarm} scalability test on GitLab CI.}
    \label{breakdown-gl}
\end{figure}

\textbf{Fig. \ref{breakdown}} and \textbf{Fig. \ref{breakdown-gl}} show the time breakdown of CI for FleCSALE with \texttt{BeeSwarm} scalability test, including the original correctness test on Travis CI and one set of multi-node scalability tests using \texttt{BeeSwarm}. The scalability test involves different execution configurations that range from 1 process to 128 processes. We can see the major overhead of \texttt{BeeSwarm} comes from deploying the scalability test environment. This is mainly caused by long instance launching time on Chameleon cloud. However, since CI tests are usually not on the critical path of applications' development process, the extra time cost brings negligible impact to developers.   

\subsubsection{Scalability of \texttt{BeeSwarm} } 
\begin{figure}[h]
    \centering
    \includegraphics[width=0.48\textwidth]{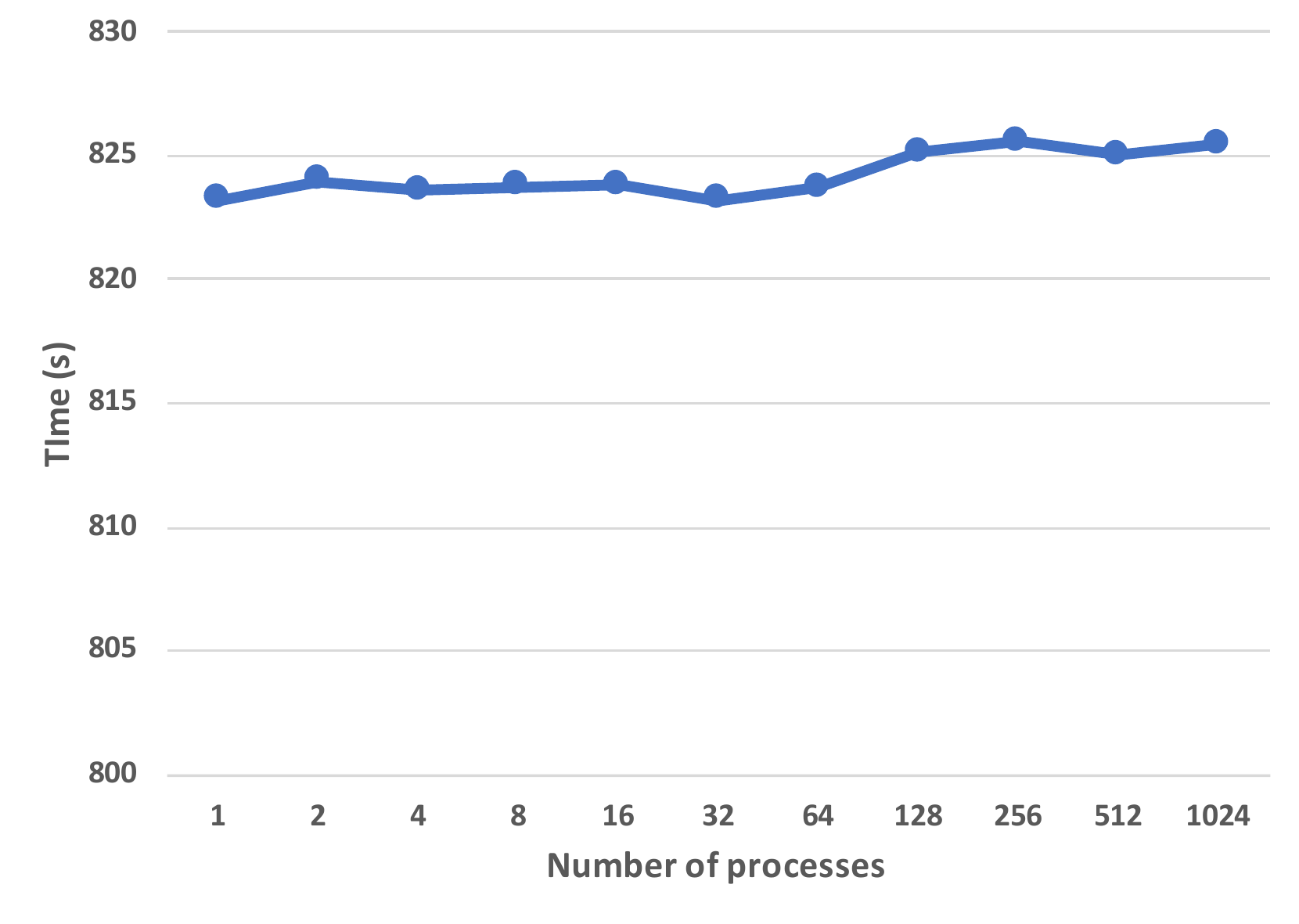}
    \caption{Scalability of deploying scalability test environment for \texttt{BeeSwarm} on Chameleon cloud.}
    \label{scalability}
\end{figure}

Since BeeSwarm is designed for launching large-scaled parallel applications, the scalability of \texttt{BeeSwarm} itself is also very important. 
As we mentioned before, the main overhead of \texttt{BeeSwarm} comes from deploying the scalability test environment. \textbf{Fig. \ref{scalability}} shows the performance of deploying the scalability test environment for \texttt{BeeSwarm}. We test it with an increasing number of processes ranging from 1 to 1024. 
We run the scalability test on 16 instances on Chameleon cloud. Each instance has 64 cores. From \textbf{Fig. \ref{scalability}}, we can see the time cost is nearly constant (less than 900 seconds) as we increase the number of process. This indicate the scalability of \texttt{BeeSwarm} itself is sufficient for large-scale test.


\subsection{Scalability Test Showcase}

\begin{figure}[h]
    \centering
    \includegraphics[width=0.45\textwidth]{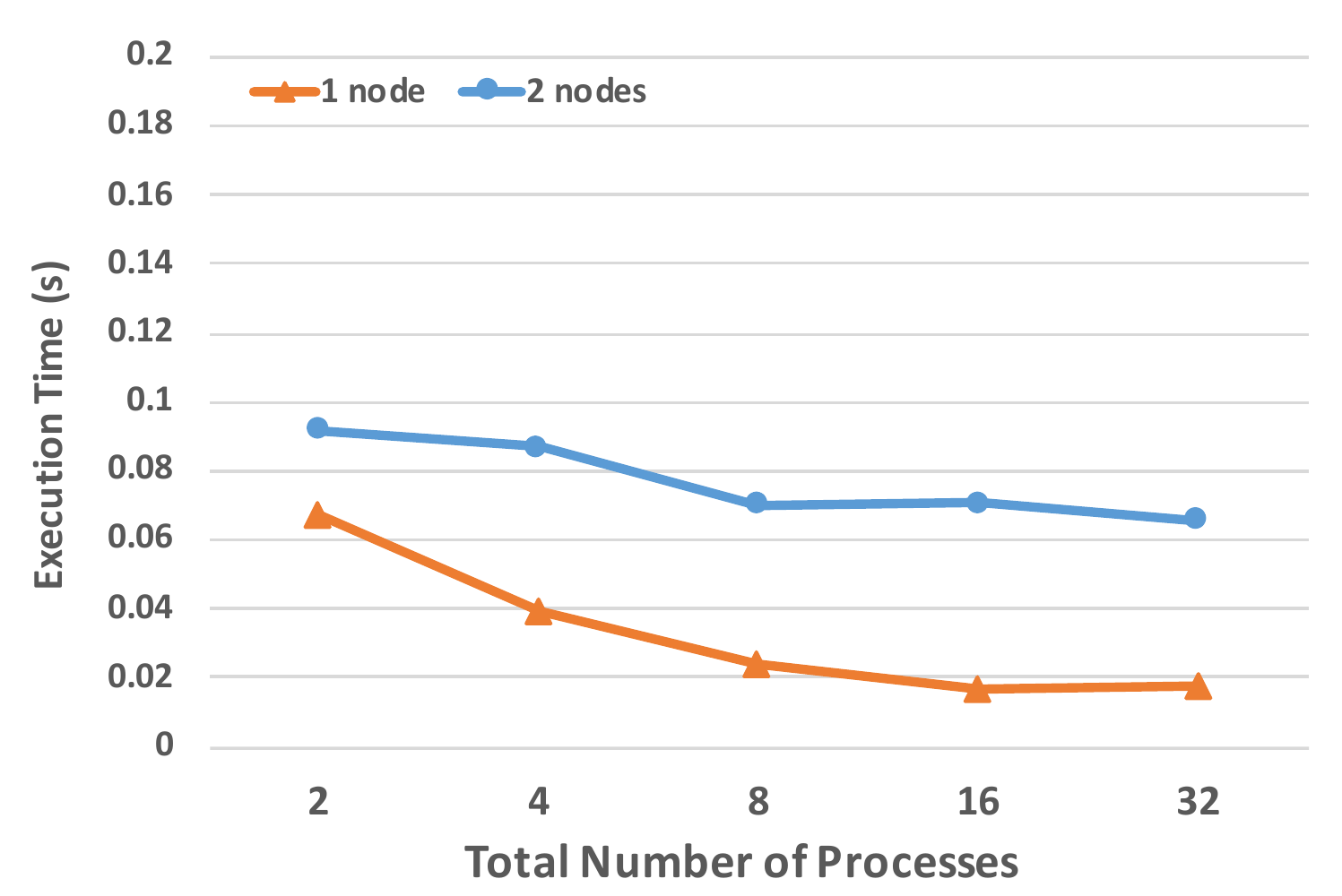}
    \caption{The scalability test result of FleCSALE.}
    \label{flecsale-result}
\end{figure}

We use FleCSALE to showcase a sample scalability test using \texttt{BeeSwarm}. We configure it to run using 2 to 32 processes on one or two nodes. When using two nodes, we evenly divide the total number of processes among them (each has 1 to 16 processes). The file generated by \texttt{BeeSwarm} is in the comma separated values (CSV) file format, and we plotted the result data in \textbf{Fig. \ref{flecsale-result}}. 
Even with this simple test using \texttt{BeeSwarm}, we can observe some interesting behavior of FleCSALE. We can see FleCSALE gains better speedup (1.73x - 4.01x) on a single node environment compared to the speedup on two nodes (1.05x - 1.40x) given the same total number of processes. This may suggest that inter-node communication could be a performance bottleneck for FleCSALE running on systems similar to Chameleon.


This result can effectively give developers the scalability data of the application they are developing, so that they can make adjustment to their application in a more timely manner. Not only can the developer observe behavior of different processing schemes, but using \texttt{BeeSwarm} can help aid them to see performance improvement or degradation of their application as they push changes to the application. 
\section{Related Work}
\label{related_work}
Scalability is one of the most important metric when we evaluate the quality of HPC applications. Many works have been done to build scalability test tools to facilitate HPC application development. For example, \cite{vetter2005mpip} proposed a a lightweight profiling library for MPI applications, which is only based on statistical information about MPI functions and brings little performance overhead. \cite{chen2006stas} proposed a effective scalability testing and analysis system -- STAS. \cite{chung2006mpi} proposed a configurable MPI scalability analysis tool for Blue Gene/L supercomputer. \cite{brunst2013custom} proposed a performance tool, Vampir, that can be used to detect hot spots in HPC applications. This can efficiently help HPC developers make their applications more scalable. \cite{merchant2012tool} proposed JACE (Job Auto-creator and Executor), a tool that enables automation of creation and execution of complex performance and scalability regression tests. It can help developers tune an application on a given platform to maximize performance given different optimization flags and tunable variables. \cite{muraleedharan2012hawk} presented a HPC performance and scalability test tool, Hawk-i, that uses cloud computing platforms to test HPC applications in order to reduce the effort to access relative scarce and on-demand high performance resources. \cite{bell2003paraprof} proposed, ParaProf, a portable, extensible, and scalable tool for parallel performance profile analysis. It gathers rich number of hardware counters and traceable information in order to offer much more detailed profiling result similar to state-of-the-art single process profiling tools. \cite{yoo2015patha} proposed a scalability test tool, PATHA, that uses system logs to extract key performance measures and apply the statistical tools and data mining methods on the performance data to identify bottlenecks or to debug the performance issues in HPC applications. Although these recent work is useful in scalability test for HPC applications, their tools or systems cannot be easily adopted by current HPC application development projects since they either require modification to the HPC application or a complicated installation or configuration process in order to make their tools working properly on a given HPC platform. 

\section{Conclusion}
\label{conclusion}
In this work, we first discuss the benefit of CI in the software development process. Then, we propose to bring scalability tests to CI so that developers can also get feedback about their applications in terms of scalability in addition to functionality. We design \texttt{BeeSwarm}, as a scalability test system for most CI environments. It is easy to use and can be integrated into any software development workflow. A variety of computing platforms can be used as a computing back-end for scalability tests. Experiments were conducted on Travis CI and GitLab CI with Chameleon Cloud as computing backend. Experimental results show that \texttt{BeeSwarm} offers good performance and scalability on large scale tests. 

\bibliographystyle{IEEEtran}
\bibliography{biblio}

\begin{thebibliography}{10}
\providecommand{\url}[1]{#1}
\csname url@samestyle\endcsname
\providecommand{\newblock}{\relax}
\providecommand{\bibinfo}[2]{#2}
\providecommand{\BIBentrySTDinterwordspacing}{\spaceskip=0pt\relax}
\providecommand{\BIBentryALTinterwordstretchfactor}{4}
\providecommand{\BIBentryALTinterwordspacing}{\spaceskip=\fontdimen2\font plus
\BIBentryALTinterwordstretchfactor\fontdimen3\font minus
  \fontdimen4\font\relax}
\providecommand{\BIBforeignlanguage}[2]{{%
\expandafter\ifx\csname l@#1\endcsname\relax
\typeout{** WARNING: IEEEtran.bst: No hyphenation pattern has been}%
\typeout{** loaded for the language `#1'. Using the pattern for}%
\typeout{** the default language instead.}%
\else
\language=\csname l@#1\endcsname
\fi
#2}}
\providecommand{\BIBdecl}{\relax}
\BIBdecl

\bibitem{fowler2006continuous}
M.~Fowler and M.~Foemmel, ``Continuous integration,'' \emph{Thought-Works)
  http://www. thoughtworks. com/Continuous Integration. pdf}, vol. 122, p.~14,
  2006.

\bibitem{bauer2012legion}
M.~Bauer, S.~Treichler, E.~Slaughter, and A.~Aiken, ``Legion: Expressing
  locality and independence with logical regions,'' in \emph{Proceedings of the
  international conference on high performance computing, networking, storage
  and analysis}.\hskip 1em plus 0.5em minus 0.4em\relax IEEE Computer Society
  Press, 2012, p.~66.

\bibitem{ferenbaugh2015pennant}
C.~R. Ferenbaugh, ``Pennant: an unstructured mesh mini-app for advanced
  architecture research,'' \emph{Concurrency and Computation: Practice and
  Experience}, vol.~27, no.~17, pp. 4555--4572, 2015.

\bibitem{bee}
\BIBentryALTinterwordspacing
J.~Chen, Q.~Guan, X.~Liang, L.~J. Vernon, A.~McPherson, L.-T. Lo, Z.~Chen, and
  J.~P. Ahrens. (2017) Docker-enabled build and execution environment (bee): an
  encapsulated environment enabling hpc applications running everywhere.
  [Online]. Available: \url{arXiv:1712.06790}
\BIBentrySTDinterwordspacing

\bibitem{beeflow}
J.~Chen, Q.~Guan, Z.~Zhang, X.~Liang, L.~Vernon, A.~McPherson, L.-T. Lo,
  P.~Grubel, T.~Randles, Z.~Chen, and J.~Ahrens, ``Beeflow: a workflow
  management system for in situ processing across hpc and cloud systems,'' in
  \emph{38th IEEE International Conference on Distributed Computing Systems
  (ICDCS)}, 2018.

\bibitem{chen2018build}
J.~Chen, Q.~Guan, X.~Liang, P.~Bryant, P.~Grubel, A.~McPherson, L.-T. Lo,
  T.~Randles, Z.~Chen, and J.~P. Ahrens, ``Build and execution environment
  (bee): an encapsulated environment enabling hpc applications running
  everywhere,'' in \emph{2018 IEEE International Conference on Big Data (Big
  Data)}.\hskip 1em plus 0.5em minus 0.4em\relax IEEE, 2018, pp. 1737--1746.

\bibitem{daniel2016lanl}
D.~J. Daniel and A.~L. Hungerford, ``Lanl asc advanced technology development
  and mitigation: Next-generation code project (ngc),'' Los Alamos National
  Lab.(LANL), Los Alamos, NM (United States), Tech. Rep., 2016.

\bibitem{mambretti2015next}
J.~Mambretti, J.~Chen, and F.~Yeh, ``Next generation clouds, the chameleon
  cloud testbed, and software defined networking (sdn),'' in \emph{Cloud
  Computing Research and Innovation (ICCCRI), 2015 International Conference
  on}.\hskip 1em plus 0.5em minus 0.4em\relax IEEE, 2015, pp. 73--79.

\bibitem{charest2017flexible}
M.~R.~J. Charest, ``Flexible computational science infrastructure (flecsi):
  Overview \& application progress,'' Los Alamos National Lab.(LANL), Los
  Alamos, NM (United States), Tech. Rep., 2017.

\bibitem{vetter2005mpip}
J.~Vetter and C.~Chambreau, ``mpip: Lightweight, scalable mpi profiling,''
  2005.

\bibitem{chen2006stas}
Y.~Chen and X.-H. Sun, ``Stas: A scalability testing and analysis system,'' in
  \emph{Cluster Computing, 2006 IEEE International Conference on}.\hskip 1em
  plus 0.5em minus 0.4em\relax IEEE, 2006, pp. 1--10.

\bibitem{chung2006mpi}
I.-H. Chung, R.~E. Walkup, H.-F. Wen, and H.~Yu, ``Mpi performance analysis
  tools on blue gene/l,'' in \emph{SC 2006 Conference, Proceedings of the
  ACM/IEEE}.\hskip 1em plus 0.5em minus 0.4em\relax IEEE, 2006, pp. 16--16.

\bibitem{brunst2013custom}
H.~Brunst and M.~Weber, ``Custom hot spot analysis of hpc software with the
  vampir performance tool suite,'' in \emph{Tools for High Performance
  Computing 2012}.\hskip 1em plus 0.5em minus 0.4em\relax Springer, 2013, pp.
  95--114.

\bibitem{merchant2012tool}
S.~Merchant and G.~Prabhakar, ``Tool for performance tuning and regression
  analyses of hpc systems and applications,'' in \emph{High Performance
  Computing (HiPC), 2012 19th International Conference on}.\hskip 1em plus
  0.5em minus 0.4em\relax IEEE, 2012, pp. 1--6.

\bibitem{muraleedharan2012hawk}
V.~Muraleedharan, ``Hawk-i hpc cloud benchmark tool,'' \emph{Msc in high
  performance computing, University of Edinburgh, Edinburgh}, 2012.

\bibitem{bell2003paraprof}
R.~Bell, A.~D. Malony, and S.~Shende, ``Paraprof: A portable, extensible, and
  scalable tool for parallel performance profile analysis,'' in \emph{European
  Conference on Parallel Processing}.\hskip 1em plus 0.5em minus 0.4em\relax
  Springer, 2003, pp. 17--26.

\bibitem{yoo2015patha}
W.~Yoo, M.~Koo, Y.~Cao, A.~Sim, P.~Nugent, and K.~Wu, ``Patha: Performance
  analysis tool for hpc applications,'' in \emph{Computing and Communications
  Conference (IPCCC), 2015 IEEE 34th International Performance}.\hskip 1em plus
  0.5em minus 0.4em\relax IEEE, 2015, pp. 1--8.

\end{thebibliography}
\end{document}